\documentclass[pdflatex,sn-nature]{sn-jnl}

\usepackage{graphicx}
\usepackage{multirow}
\usepackage{amsmath,amssymb,amsfonts}
\usepackage{booktabs}
\usepackage[table]{xcolor}
\usepackage{algorithm}
\usepackage{algorithmicx}
\usepackage{algpseudocode}
\usepackage{enumitem}
\usepackage{url}
\usepackage{array}
\usepackage{rotating}
\usepackage{caption}
\usepackage{placeins}
\usepackage{tikz}
\setcitestyle{super,open={},close={}} 
\usetikzlibrary{arrows.meta,positioning,fit,calc,shapes.geometric}
\definecolor{CardinalJoint}{HTML}{E8F0F5}
\definecolor{CardinalMoE}{HTML}{E5F2F0}

\newcommand{\jointrow}{\rowcolor{CardinalJoint}}
\newcommand{\moerow}{\rowcolor{CardinalMoE}}

\title[CARDINAL for MACE risk prediction]{CARDINAL Predicts Cardiovascular Risk From Non-contrast Cardiac CT}

\author[1]{\fnm{Roy} \sur{Gabriel}}
\author[1]{\fnm{Nattakorn} \sur{Kittisut}}
\author[1]{\fnm{Jamshid} \sur{Hassanpour}}
\author[1]{\fnm{Michael} \sur{Galarnyk}}
\author[1]{\fnm{Abanoub} \sur{Abdelmalak}}
\author[2]{\fnm{Marly} \sur{van Assen}}
\author[2]{\fnm{Carlo N.} \sur{De Cecco}}
\author[2]{\fnm{Arshed} \sur{Quyyumi}}
\author*[1]{\fnm{Ali} \sur{Adibi}}\email{ali.adibi@ece.gatech.edu}

\affil*[1]{%
  \orgname{Georgia Institute of Technology},
  \orgaddress{
    \city{Atlanta},
    \state{Georgia},
    \country{USA}
  }
}

\affil[2]{%
  \orgname{Emory University Hospital},
  \orgaddress{
    \city{Atlanta},
    \state{Georgia},
    \country{USA}
  }
}

\abstract{Cardiovascular risk prediction remains limited by incomplete clinical data and imaging biomarkers that reduce computed tomography (CT) to a small number of handcrafted features. We developed CARDINAL (Cardiovascular Assessment via Representation learning from Deep Imaging with Nested Anatomical Latent embeddings), a clinically grounded framework that learns compact representations from routine non-contrast cardiac CT for major adverse cardiovascular event (MACE) prediction. In 17,659 patients, CARDINAL was evaluated for 1-, 3-, 5-, and 10-year MACE prediction against American Heart Association (AHA) pooled cohort equations (PCE), AHA predicting risk of cardiovascular disease events (PREVENT), coronary artery calcium (CAC), segmentation-derived CT biomarkers, and 70-feature structural radiomics. Gains were largest at longer horizons. At 10 years, CARDINAL (joint) achieved an area under the receiver operating characteristic curve (AUROC) of 0.866 $\pm$ 0.020 and an area under the precision--recall curve (AUPRC) of 0.890 $\pm$ 0.015, compared with an AUROC of 0.826 $\pm$ 0.023 and an AUPRC of 0.826 $\pm$ 0.022 for structural radiomics, the strongest baseline. CARDINAL also achieved the highest survival concordance index (C-index), 0.753 $\pm$ 0.015, and high-versus-low risk-tertile hazard ratio, 10.78 $\pm$ 3.16, with favorable reclassification and exploratory calibration. These findings suggest that non-contrast cardiac CT contains prognostic information beyond conventional risk equations, CAC scoring, and engineered imaging biomarkers.}

\graphicspath{{../Figures/}{../Supplementary_Information/Figures/}}

\begin{document}
\maketitle



\section{Introduction}

Cardiovascular disease remains the leading cause of death worldwide and continues to impose a major burden across health systems despite decades of progress in prevention, diagnosis, and treatment \cite{vaduganathan2022global}. Identifying individuals at high risk before the occurrence of major adverse cardiovascular events (MACE) remains a central challenge in cardiovascular medicine. In clinical practice, risk estimation guides decisions regarding statin therapy, blood pressure control, laboratory monitoring, and follow-up care \cite{Arnett2019-hr, aha-it}. However, current risk stratification strategies remain imperfect, particularly in heterogeneous real-world populations; reported calibration differences in multiethnic cohorts motivated descriptive demographic subgroup evaluation \cite{defilippis2015calibration,rana2016accuracy}.

In the United States, current preventive care relies heavily on structured clinical risk equations, most prominently the American Heart Association pooled cohort equations (PCE) and, more recently, the PREVENT (predicting risk of cardiovascular disease events) equations \cite{khan2024development, goff2014american}. These models are well validated and widely used, but they depend on clinical variables such as blood pressure, lipids, smoking status and kidney function that may be missing, outdated, or not temporally aligned with clinical decision-making. More fundamentally, these tools estimate risk from population-level associations rather than directly measuring disease burden in an individual patient. As a result, these models may not fully reflect an individual patient’s underlying disease burden or patterns of cardiovascular remodeling. Because treatment decisions are often based on risk thresholds, inaccuracies can lead to meaningful differences in clinical management \cite{diao2024projected,cho2025aha,li2022automating}.

Imaging provides a complementary approach by directly capturing structural manifestations of disease. Among imaging biomarkers, coronary artery calcium (CAC) derived from non-contrast cardiac computed tomography (CT) is the most clinically established, with strong evidence linking calcified plaque burden to future cardiovascular events across populations and follow-up intervals \cite{agatston1990quantification,detrano2008coronary,budoff2018tenyear,mcclelland2015tenyear,nasir2015implications,Hecht2017-wc,thanassoulis2012genetic,grundy20192018}. Advances in automated CAC scoring have enabled reliable extraction of calcium-based risk information from CT at scale \cite{Eng2021-ub, zeleznik2021deep, Martin2020-df, Winkel2022-ml}. These methods, along with broader advances in medical image analysis \cite{hosny2018artificial, hosny2018deep, de2018clinically, litjens2017survey, wasserthal2023totalsegmentator, Follmer2024-zh}, have expanded the ability to quantify anatomical features from CT imaging. For example, cardiac chamber volumes and thoracic anatomy have well-established associations with mortality and cardiovascular outcomes \cite{arsanjani2014left, miller2024predicting, marcinkiewicz2025holistic, ambale2017left, vukadinovic2023deep, naghavi2024ai, naghavi2024artificial}, and their relationship to aortic morphology has also been explored in the context of atherosclerotic disease \cite{DaneilJACC}.

Despite its clinical utility, CAC remains a scalar summary of disease burden. It reflects calcified coronary plaque but does not capture the broader cardiothoracic and vascular information present in CT imaging. To address this limitation, recent work has explored more comprehensive CT-derived biomarkers. In particular, structural radiomics approaches that quantify cardiac and vascular morphology have demonstrated improved prediction beyond clinical risk equations and provide complementary value when combined with CAC, although radiomics alone does not consistently outperform CAC as a standalone biomarker \cite{DaneilJACC}. However, these approaches remain dependent on predefined features derived from segmentation pipelines and may therefore miss broader or more complex patterns present in the image.

More generally, most CT-based cardiovascular prediction pipelines follow a multi-step paradigm in which anatomical segmentation and predefined feature extraction are followed by downstream modeling. While interpretable, this approach is inherently limited by the choice of engineered features and may not fully leverage the information contained in the full CT volume. An alternative is to learn representations directly from imaging data, where features are derived from the image itself rather than predefined by design. In this framework, the learned representation can be used as input to downstream prediction models, combining the flexibility of data-driven feature learning with the practicality of structured modeling. The key challenge is ensuring that such representations remain clinically meaningful and interpretable, rather than purely data-driven abstractions.

In this paper, we present a different approach that learns a clinically grounded representation directly from electrocardiogram-gated non-contrast cardiac CT. To achieve this goal, we developed CARDINAL (Cardiovascular Assessment via Representation learning from Deep Imaging with Nested Anatomical Latent embeddings), a framework that compresses the full CT volume into a low-dimensional latent representation while preserving clinically meaningful information. CARDINAL used Matryoshka Representation Learning (MRL) \cite{kusupati2024matryoshkarepresentationlearning} to produce a nested family of embeddings in which progressively larger representations retain progressively more information. To ensure clinical relevance, the representation was supervised using anatomy- and calcium-related CT-derived phenotypes.

We evaluated CARDINAL as a strictly unimodal imaging framework using non-contrast cardiac CT. A joint encoder learned one global representation, whereas a mixture-of-experts (MoE) model fused organ-specific representations. The frozen embeddings were evaluated for MACE classification across 1-, 3-, 5-, and 10-year horizons and for time-to-event modeling. 


Our final results show that such a representation can improve cardiovascular risk prediction and stratification by demonstrating that CARDINAL: (1) preserves clinically meaningful anatomical and calcium-related information within its latent space; (2) improves discrimination of MACE across multiple time horizons relative to PCE, PREVENT, CAC, and radiomics-based CT biomarkers; and (3) improves survival discrimination and patient-level risk stratification. Through these analyses, we establish clinically grounded latent representation learning as a more effective paradigm for cardiovascular risk prediction, showing that it can capture prognostic information beyond what is accessible through conventional clinical scores or engineered imaging features.

\section{Results}
\subsection{Study population}
This retrospective study contained 28,092 CAC CT examinations using standard protocol \cite{Hecht2017-wc} from 25,514 adults acquired between 2010 and 2023 across 11 Emory Healthcare-affiliated sites. Examinations were acquired using scanners from five manufacturers: Philips, Siemens, General Electric, Toshiba, and Canon Medical Systems. History of atherosclerotic cardiovascular disease (ASCVD) and imaging indication were determined using historical International Classification of Diseases (ICD) and Current Procedural Terminology (CPT) codes together with CT order information. The study received Emory University Institutional Review Board approval and a waiver of informed consent. Electronic health record (EHR) variables used for clinical risk estimation were obtained at or within 6 months before imaging.

MACE were defined as stroke, myocardial infarction, percutaneous coronary intervention, coronary artery bypass grafting occurring more than 90 days after CT, or all-cause mortality. To validate outcome definitions, 10\% of the cohort were manually reviewed including at least 10\% representation from each event category and the non-event group, confirming the accuracy of code-based event classification.

Two horizon-labeling strategies were evaluated. In the \textit{withFU} (accounting for follow-up) setting, only patients with documented follow-up through the specified time horizon or an earlier event were included, yielding a smaller but more rigorously observed cohort. In the \textit{ignoreFU} (not accounting for follow-up) setting, absence of a recorded event by the specified time horizon was treated as a negative label regardless of follow-up completeness, resulting in a larger cohort that better reflects the overall imaged population but may underestimate event prevalence.

\begin{figure*}[!tb]
\centering
\includegraphics[width=0.98\textwidth]{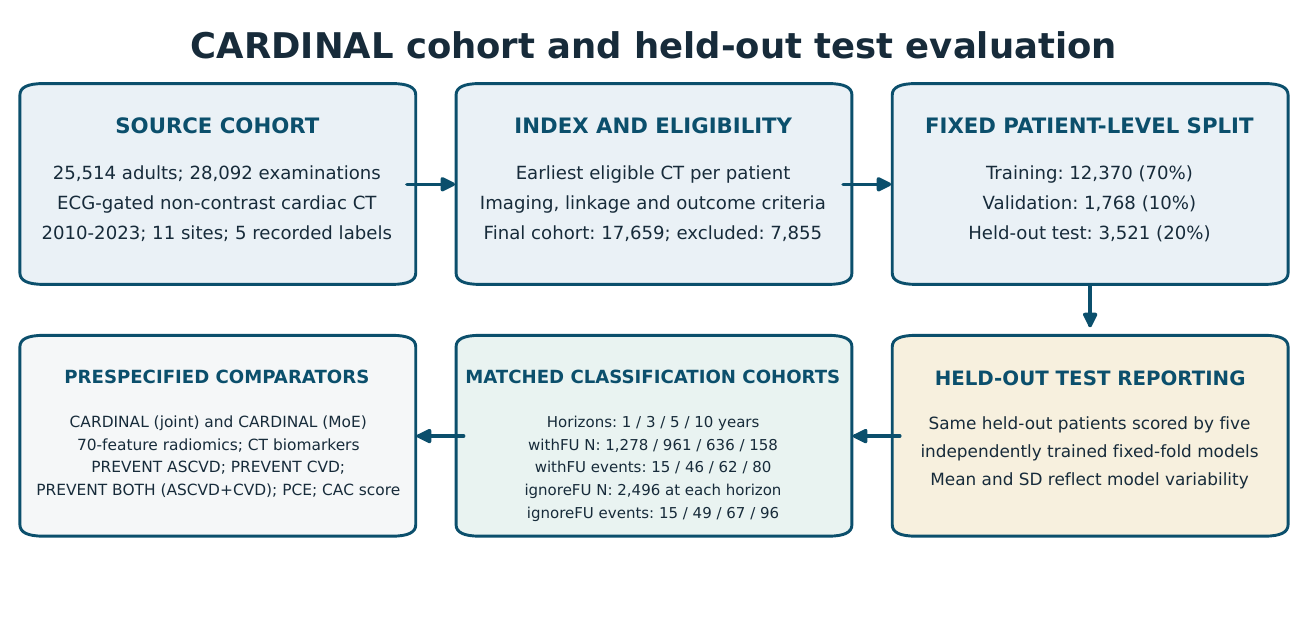}
\caption{\textbf{Patient flow and held-out test evaluation.} The diagram summarizes source examinations, index-examination selection, cohort eligibility, the fixed patient-level split and horizon-specific matched test cohorts. The same held-out patients were evaluated by five independently trained fixed-fold models.}
\label{fig:consort}
\end{figure*}

After retaining the earliest eligible CT (index CT) per patient and applying imaging, preprocessing, outcome, and data-linkage eligibility criteria, 17,659 patients remained. The fixed patient-level partition contained 12,370 training, 1,768 validation, and 3,521 held-out test patients (Fig.~\ref{fig:consort}).

In the source clinical dataset, a subset of patients underwent multiple CT examinations. Specifically, 2,369 of 25,514 patients (9.3\%) had more than one CT scan, with a maximum of five scans per patient. To ensure statistical independence and prevent information leakage across dataset splits, the final analysis cohort was restricted to a single index CT per patient, defined as the earliest eligible examination. Consequently, each patient contributed only one observation to model development and evaluation.

In the held-out test partition, mean age was 56.0 $\pm$ 10.0 years and 59.9\% were male; 77.5\% were White, 8.6\% Black or African American, and 3.7\% Asian. Median follow-up was 4.02 years (interquartile range [IQR], 2.24--6.19). Across the full analytic cohort, the 10-year event rate was 3.18\% (561/17,659) under \textit{ignoreFU} and 59.18\% (493/833) under \textit{withFU}. Baseline characteristics and horizon-specific outcome counts are presented together in Table~\ref{tab:cohort}. Absolute standardized mean differences across train, validation, and test partitions were all below 0.10 (maximum, 0.098).

\begin{table*}[!h]
\caption{\textbf{Baseline characteristics and outcome frequency in the fixed patient-level cohort.} Continuous variables are mean $\pm$ standard deviation (SD) among available observations unless otherwise stated. Categorical variables are n (\%). Coronary artery calcium (CAC) score is shown as median (IQR) and mean $\pm$ SD because of its zero-inflated, right-skewed distribution.}
\label{tab:cohort}
\centering\footnotesize
\renewcommand{\arraystretch}{0.94}
\setlength{\tabcolsep}{4pt}
\begin{tabular}{@{}>{\raggedright\arraybackslash}p{0.39\textwidth}>{\centering\arraybackslash}p{0.18\textwidth}>{\centering\arraybackslash}p{0.18\textwidth}>{\centering\arraybackslash}p{0.18\textwidth}@{}}\toprule
\multicolumn{4}{@{}l}{\textbf{a, Baseline characteristics}}\\
 & \textbf{Training} & \textbf{Validation} & \textbf{Held-out test} \\
N patients & 12,370 & 1,768 & 3,521 \\
Age, years & 55.8 $\pm$ 10.0 & 56.0 $\pm$ 9.9 & 56.0 $\pm$ 10.0 \\
Female sex, n (\%) & 5062 (40.9) & 728 (41.2) & 1412 (40.1) \\
Male sex, n (\%) & 7308 (59.1) & 1040 (58.8) & 2109 (59.9) \\
Asian race, n (\%) & 443 (3.6) & 69 (3.9) & 131 (3.7) \\
Black or African American race, n (\%) & 1179 (9.5) & 145 (8.2) & 303 (8.6) \\
Other race, n (\%) & 1224 (9.9) & 170 (9.6) & 358 (10.2) \\
White race, n (\%) & 9524 (77.0) & 1384 (78.3) & 2729 (77.5) \\
Body mass index (BMI), kg/m$^2$ & 27.9 $\pm$ 5.4 & 28.0 $\pm$ 5.4 & 27.9 $\pm$ 5.4 \\
Hypertension, n (\%) & 5615 (45.4) & 779 (44.1) & 1681 (47.7) \\
Diabetes mellitus, n (\%) & 1320 (10.7) & 184 (10.4) & 367 (10.4) \\
Systolic blood pressure, mmHg & 126.8 $\pm$ 17.9 & 126.9 $\pm$ 18.8 & 128.5 $\pm$ 17.8 \\
Total cholesterol, mg/dL & 195.8 $\pm$ 43.5 & 195.1 $\pm$ 42.3 & 195.9 $\pm$ 43.4 \\
High-density lipoprotein (HDL) cholesterol, mg/dL & 54.9 $\pm$ 16.9 & 55.1 $\pm$ 16.4 & 54.6 $\pm$ 16.3 \\
Estimated glomerular filtration rate (eGFR), mL/min/1.73 m$^2$ & 78.0 $\pm$ 20.9 & 79.0 $\pm$ 20.5 & 77.9 $\pm$ 21.5 \\
CAC score, median (IQR) & 2 (0--56) & 3 (0--66) & 2 (0--56) \\
CAC score, mean $\pm$ SD & 105.4 $\pm$ 321.8 & 119.8 $\pm$ 334.6 & 110.9 $\pm$ 352.7 \\
\bottomrule
\end{tabular}
\vspace{6pt}

\begin{tabular}{@{}lcc@{}}\toprule
\multicolumn{3}{@{}l}{\textbf{b, MACE outcome counts by horizon}}\\
Outcome & \textit{ignoreFU} & \textit{withFU} \\
1-year & 94/17,659 (0.53\%) & 92/6,899 (1.33\%) \\
3-year & 256/17,659 (1.45\%) & 237/5,089 (4.66\%) \\
5-year & 393/17,659 (2.23\%) & 358/3,227 (11.09\%) \\
10-year & 561/17,659 (3.18\%) & 493/833 (59.18\%) \\
\bottomrule
\end{tabular}
\end{table*}


\subsection{Clinical and imaging baselines}
We compared CARDINAL with PCE, PREVENT ASCVD, PREVENT CVD, both PREVENT scores (PREVENT BOTH), CAC score, six segmentation-derived CT biomarkers, and a distinct 70-feature structural radiomics benchmark. Both PCE and PREVENT were computed from EHR variables recorded at or within 6 months before imaging and the analyses used complete required inputs without imputation. The six CT biomarkers were aorta, heart, lung, and myocardium volumes, CAC volume, and CAC score. The additional radiomics benchmark comprised 14 three-dimensional shape features from each of five cardiac and vascular structures \cite{wasserthal2023totalsegmentator,DaneilJACC,vanGriethuysen2017pyradiomics,zwanenburg2020ibsi}. Since PCE has been widely replaced by PREVENT, PCE results are retained in Supplementary Information.

All displayed models were evaluated on the same held-out patients within each horizon and labeling convention. Together, these comparators span established clinical risk equations, scalar imaging biomarkers, and engineered CT-derived features.


\subsection{Clinically grounded latent representations}
We first asked whether CARDINAL preserved clinically interpretable CT-derived phenotypes across nested latent dimensions. Representation learning was supervised using six anatomy- and calcium-grounded auxiliary targets derived directly from CT: aorta, heart, lung, and myocardium volumes, CAC volume, and CAC score. Performance was evaluated across the various Matryoshka-based depth sweep, $d\in{1,2,4,\ldots,1024}$, and is reported as mean $\pm$ standard deviation across five fixed fold models on the held-out test set. The representation-learning procedure is described in the Methods.

\begin{figure*}[!tb]
\centering
\includegraphics[width=\textwidth]{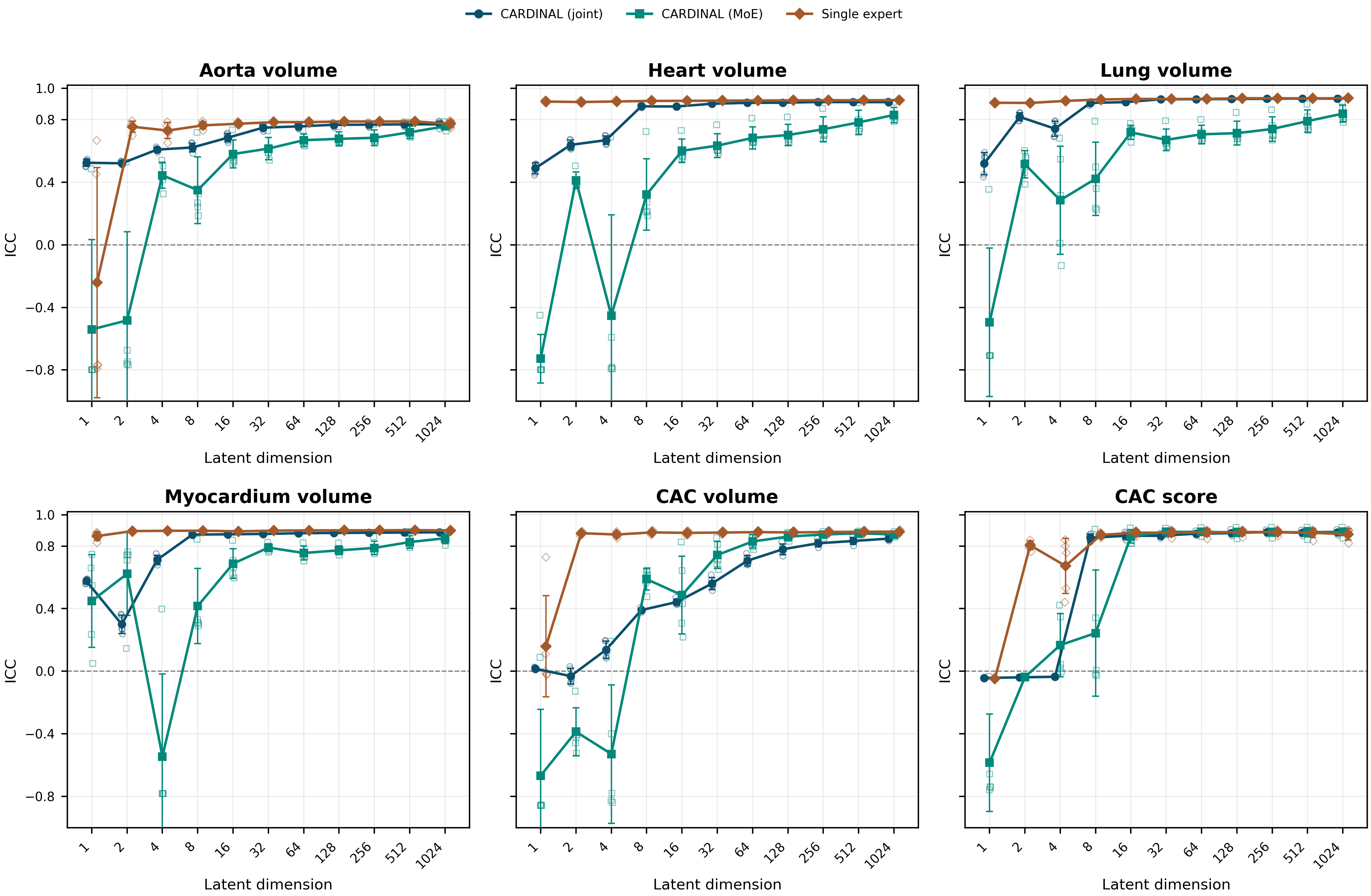}
\caption{\textbf{Phenotype recovery across nested latent dimensions.} Open symbols show fold-specific ICC values from the five fixed-fold models, solid symbols and connecting lines show the corresponding means, and vertical error bars use the sample standard deviation computed directly from the five fold-specific ICC values at each depth. All Matryoshka depths from 1 to 1024 are displayed. Complete ICC, $R^2$, and mean absolute error results are reported in Supplementary Information.}\label{fig:aux}
\end{figure*}

CARDINAL (joint), which learns a single shared representation for all six targets, retained substantial clinically meaningful information even within compact latent prefixes. At dimension of size $d=8$, mean Intraclass Correlation Coefficients (ICCs) were 0.885 $\pm$ 0.007 for heart volume, 0.907 $\pm$ 0.012 for lung volume, and 0.873 $\pm$ 0.007 for myocardium volume. Performance generally increased with latent capacity, with the highest observed CARDINAL (joint) ICCs ranging from 0.770 $\pm$ 0.020 for aorta volume to 0.935 $\pm$ 0.003 for lung volume (Fig.~\ref{fig:aux}).

The independently trained single-expert models served as task-specific references and achieved the highest ICC for five of the six phenotypes. CARDINAL (MoE) fused these expert representations into a unified latent space from which all six phenotypes were recovered simultaneously. The fused representation retained particularly strong calcium-related information, reaching an ICC of 0.882 $\pm$ 0.011 for CAC volume and 0.891 $\pm$ 0.025 for CAC score. Unlike the individual experts, the joint and MoE formulations provide unified representations that can be reused across downstream tasks. Complete depth-level ICC, $R^2$, and mean absolute error results are provided in Supplementary Information.

\subsection{MACE classification on the held-out test set}

We next evaluated downstream MACE prediction using frozen CARDINAL embeddings as image representations and compared performance against PREVENT ASCVD, PREVENT CVD, both PREVENT scores used together as downstream inputs (PREVENT BOTH), CAC score, a six-feature CT biomarker baseline, and a distinct 70-feature structural radiomics baseline. Classification was evaluated at 1-, 3-, 5-, and 10-year horizons under both \textit{withFU} and \textit{ignoreFU}.

\begin{figure*}[!tb]
\centering
\includegraphics[width=\textwidth]{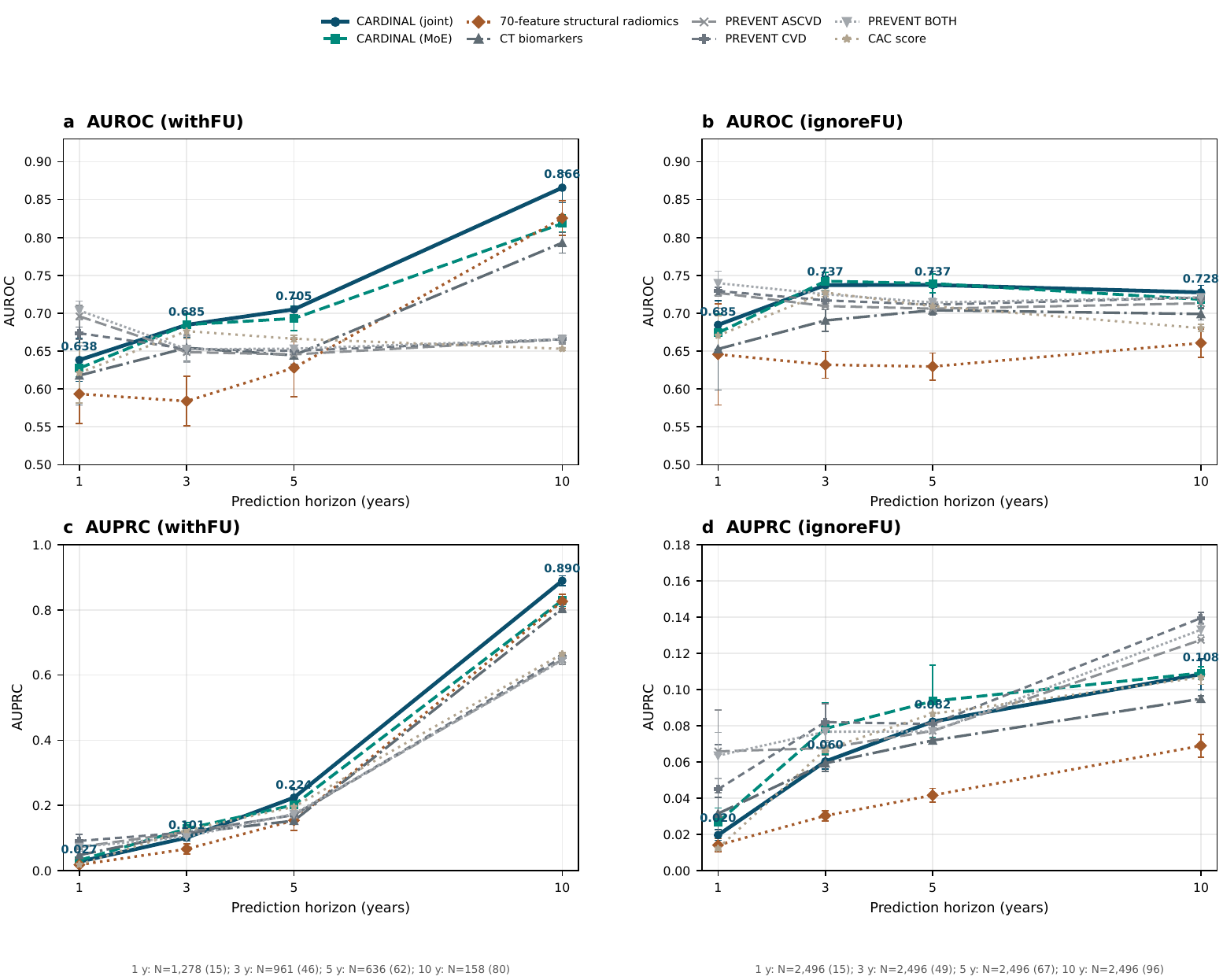}
\caption{\textbf{Held-out test classification across prediction horizons.} \textbf{a,b}, Mean AUROC across five independently trained fixed-fold models at each horizon under \textit{withFU} and \textit{ignoreFU}. \textbf{c,d}, Corresponding mean AUPRC. Vertical error bars use the sample standard deviation computed directly from the five fold-specific AUROC or AUPRC values at each horizon, and CARDINAL (joint) means are labeled directly. PREVENT ASCVD, PREVENT CVD, and PREVENT BOTH are displayed as separate clinical baselines. PREVENT BOTH uses both scores as downstream inputs. Numbers of evaluated patients and events are reported below each panel and in Supplementary Information.}\label{fig:classification}
\end{figure*}

The strongest overall classification result was achieved by CARDINAL (joint) at 10 years under \textit{withFU}. At $d=1024$, CARDINAL (joint) achieved AUROC $=0.866 \pm 0.020$ and AUPRC $=0.890 \pm 0.015$. This exceeded structural radiomics (AUROC $=0.826 \pm 0.023$, AUPRC $=0.826 \pm 0.022$), CARDINAL (MoE) (AUROC $=0.818 \pm 0.011$, AUPRC $=0.829 \pm 0.019$), CT biomarkers (AUROC $=0.793 \pm 0.014$, AUPRC $=0.804 \pm 0.012$), PREVENT ASCVD (AUROC $=0.665 \pm 0.006$, AUPRC $=0.647 \pm 0.014$), PREVENT CVD (AUROC $=0.666 \pm 0.000$, AUPRC $=0.655 \pm 0.000$), PREVENT BOTH (AUROC $=0.666 \pm 0.004$, AUPRC $=0.645 \pm 0.005$), and CAC score (AUROC $=0.653 \pm 0.002$, AUPRC $=0.665 \pm 0.005$) (Fig.~\ref{fig:classification}; Table~\ref{tab:primaryevidence}).

At shorter horizons, CARDINAL remained competitive. Under \textit{withFU}, CARDINAL (joint) and CARDINAL (MoE) tied for the highest mean AUROC at 3 years, and CARDINAL (joint) achieved the highest AUROC and AUPRC at 5 years. Under \textit{ignoreFU}, CARDINAL (MoE) achieved the highest AUROC at 3 and 5 years, while CARDINAL (joint) achieved the highest AUROC at 10 years. PREVENT BOTH achieved the highest AUROC in the 1-year analyses. Complete horizon-specific AUROC and AUPRC values are reported in Supplementary Information.

On paired patient-level predictions at 10 years under \textit{withFU}, DeLong testing showed significantly higher AUROC for CARDINAL (joint) than CAC score ($P=7.39\times10^{-10}$), PREVENT ASCVD ($P=4.67\times10^{-6}$), PREVENT CVD ($P=5.71\times10^{-6}$), PREVENT BOTH ($P=8.54\times10^{-6}$), and CT biomarkers ($P=5.76\times10^{-3}$). Comparisons with CARDINAL (MoE) ($P=0.082$) and structural radiomics ($P=0.249$) were not statistically significant. Complete model configurations, paired DeLong comparisons, and fold-level Wilcoxon tests are provided in Supplementary Information.

\subsection{Survival prediction and risk stratification}

Because cardiovascular outcomes are time-to-event processes, we complemented horizon classification with penalized Cox modeling in patients with valid event or censoring times. CARDINAL (joint) ($d=512$) achieved the highest concordance index (C-index), 0.753 $\pm$ 0.015, followed by CARDINAL (MoE) ($d=1024$), 0.733 $\pm$ 0.009. Structural radiomics and CT biomarkers achieved C-indices of 0.687 $\pm$ 0.010 and 0.672 $\pm$ 0.015, respectively (Table~\ref{tab:primaryevidence}). The exact two-sided fold-level Wilcoxon comparison between CARDINAL (joint) and structural radiomics yielded $P=0.0625$, the smallest attainable value with five non-zero paired fold differences.

CARDINAL (joint) also produced the largest high-versus-low predicted-risk tertile hazard ratio (HR), 10.78 $\pm$ 3.16, compared with 7.92 $\pm$ 1.74 for CARDINAL (MoE), 5.92 $\pm$ 1.30 for structural radiomics, 4.71 $\pm$ 1.31 for CT biomarkers, 2.92 $\pm$ 0.00 for CAC score, and 2.82 $\pm$ 0.00 for PREVENT ASCVD. Both CARDINAL variants separated low-, intermediate-, and high-risk groups by log-rank testing ($P<0.0001$; Fig.~\ref{fig:latent_survival}). Proportional-hazards diagnostics and complete fold-level survival outputs are reported in Supplementary Information.

\begin{table*}[!tb]
\caption{\textbf{Primary 10-year classification evidence and survival performance.}
\textbf{a}, Classification metrics were evaluated on the globally matched
10-year \textit{withFU} held-out cohort. AUROC and
AUPRC are mean $\pm$ sample standard deviation across five independently
trained fixed-fold models. $\Delta$AUROC is CARDINAL (joint) minus the
comparator and was calculated from unrounded fold means. DeLong tests used
paired patient-level predictions. Continuous net reclassification improvement
(NRI) was also calculated from paired predictions; positive values favor
CARDINAL (joint).
\textbf{b}, Survival values are mean $\pm$ sample standard deviation across
five fixed models. $\Delta$ C-index is CARDINAL (joint) minus the comparator.
Hazard ratios (HRs) compare the high- and low-risk tertiles. Bold denotes the
highest value within each panel.}
\label{tab:primaryevidence}

\centering
\scriptsize
\setlength{\tabcolsep}{3.2pt}
\renewcommand{\arraystretch}{1.15}

\textbf{a, Paired 10-year \textit{withFU} classification comparisons}\\[3pt]

\begin{tabular*}{\textwidth}{
@{\extracolsep{\fill}}
>{\raggedright\arraybackslash}p{0.27\textwidth}
cccc
@{}}
\toprule
\textbf{Model}
& \textbf{AUROC}
& \textbf{$\Delta$AUROC}
& \textbf{AUPRC}
& \shortstack{\textbf{NRI}\\\textbf{(95\% CI)}} \\
\midrule

\jointrow
\textbf{CARDINAL (joint)}
& \textbf{0.866 $\pm$ 0.020}
& Reference
& \textbf{0.890 $\pm$ 0.015}
& Reference \\

\moerow
\textbf{CARDINAL (MoE)}
& 0.818 $\pm$ 0.011
& +0.047
& 0.829 $\pm$ 0.019
& 0.507 (0.203--0.808) \\

\shortstack[l]{70-feature structural\\radiomics}
& 0.826 $\pm$ 0.023
& +0.040
& 0.826 $\pm$ 0.022
& 0.551 (0.272--0.856) \\

CT biomarkers
& 0.793 $\pm$ 0.014
& +0.073$^{**}$
& 0.804 $\pm$ 0.012
& 0.806 (0.525--1.086) \\

PREVENT ASCVD
& 0.665 $\pm$ 0.006
& +0.200$^{***}$
& 0.647 $\pm$ 0.014
& 0.984 (0.728--1.264) \\

PREVENT CVD
& 0.666 $\pm$ 0.000
& +0.200$^{***}$
& 0.655 $\pm$ 0.000
& 0.933 (0.655--1.213) \\

PREVENT BOTH
& 0.666 $\pm$ 0.004
& +0.200$^{***}$
& 0.645 $\pm$ 0.005
& 0.908 (0.629--1.188) \\

CAC score
& 0.653 $\pm$ 0.002
& +0.213$^{***}$
& 0.665 $\pm$ 0.005
& 1.059 (0.801--1.314) \\

\bottomrule
\end{tabular*}

\vspace{9pt}

\textbf{b, Survival discrimination and risk stratification}\\[3pt]

\begin{tabular*}{\textwidth}{
@{\extracolsep{\fill}}
>{\raggedright\arraybackslash}p{0.27\textwidth}
ccc
@{}}
\toprule
\textbf{Model}
& \textbf{C-index}
& \textbf{$\Delta$ C-index}
& \shortstack{\textbf{High-versus-low}\\\textbf{tertile HR}} \\
\midrule

\jointrow
\textbf{CARDINAL (joint)}
& \textbf{0.753 $\pm$ 0.015}
& Reference
& \textbf{10.78 $\pm$ 3.16} \\

\moerow
\textbf{CARDINAL (MoE)}
& 0.733 $\pm$ 0.009
& +0.020
& 7.92 $\pm$ 1.74 \\

\shortstack[l]{70-feature structural\\radiomics}
& 0.687 $\pm$ 0.010
& +0.066$^{\dagger}$
& 5.92 $\pm$ 1.30 \\

CT biomarkers
& 0.672 $\pm$ 0.015
& +0.081
& 4.71 $\pm$ 1.31 \\

CAC score
& 0.640 $\pm$ 0.000
& +0.113
& 2.92 $\pm$ 0.00 \\

PREVENT ASCVD
& 0.622 $\pm$ 0.000
& +0.131
& 2.82 $\pm$ 0.00 \\

\bottomrule
\end{tabular*}

\vspace{4pt}

\begin{minipage}{\textwidth}
\scriptsize
\textit{Note:}
$^{*}P<0.05$, $^{**}P<0.01$, and $^{***}P<0.001$ by paired
two-sided DeLong testing of AUROC versus CARDINAL (joint).
No significance symbols are assigned to AUPRC because fold-level Wilcoxon
comparisons were based on only five paired models and are reported in
Supplementary Information.
$^{\dagger}$Exact two-sided fold-level Wilcoxon signed-rank
$P=0.0625$ for the C-index comparison between CARDINAL (joint) and
structural radiomics.
CI, confidence interval. PREVENT BOTH uses PREVENT ASCVD and PREVENT CVD
as joint downstream inputs. Full horizon-specific metrics, paired tests,
calibration results and subgroup analyses are provided in Supplementary
Information.
\end{minipage}

\end{table*}


\subsection{Compact latent representations}

\begin{figure*}[!tb]
\centering
\includegraphics[
  width=\textwidth
]{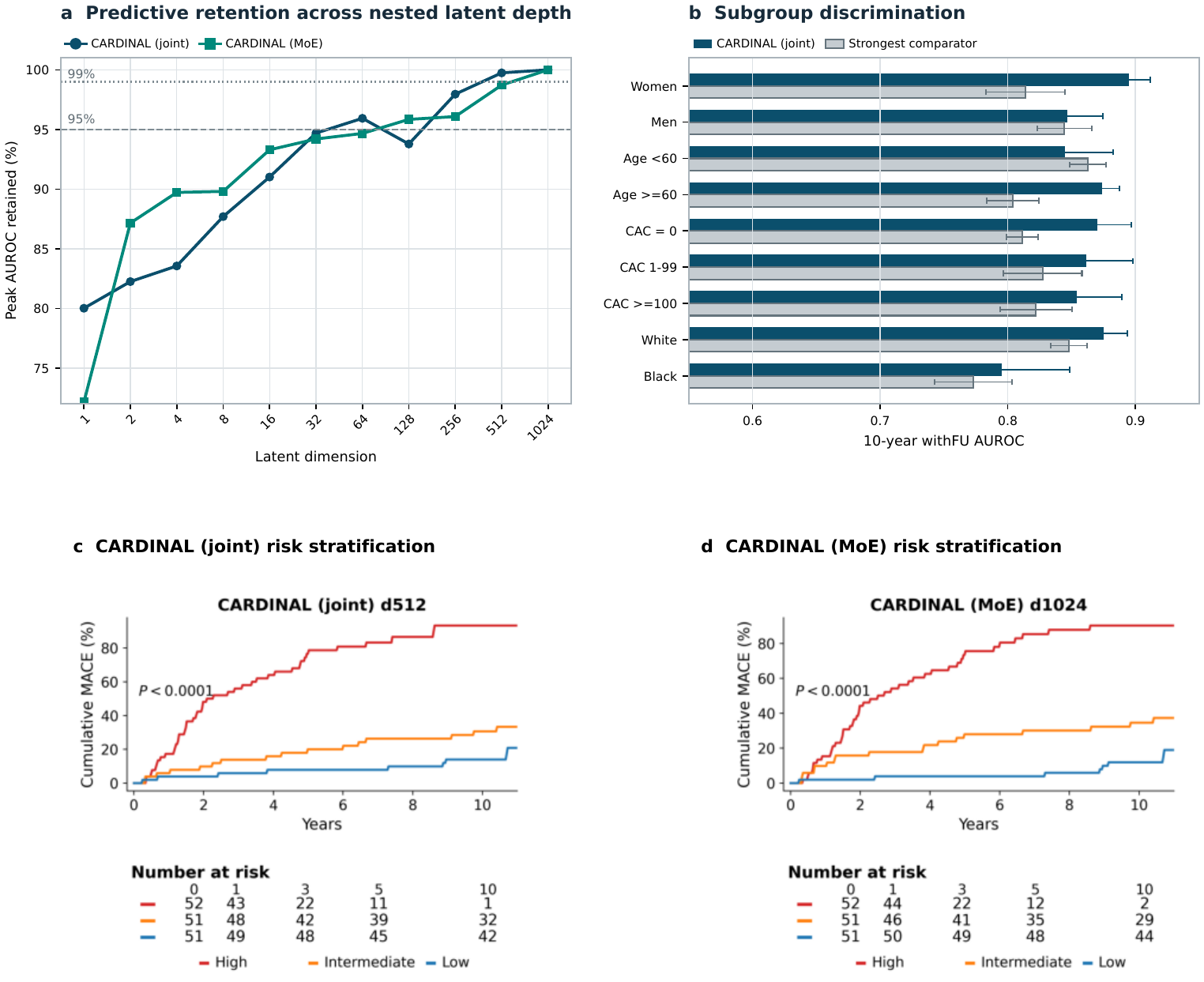}
\caption{\textbf{Compact representations, subgroup discrimination and survival risk stratification.} \textbf{a}, Percentage of peak 10-year \textit{withFU} AUROC retained across latent depths. \textbf{b}, Mean 10-year \textit{withFU} AUROC across five fixed models for CARDINAL (joint) and the strongest non-joint comparator within each sex, age, CAC, and race subgroup; error bars show the sample standard deviation across the five fixed models, and comparator identities are reported in Supplementary Information. \textbf{c,d}, Cumulative MACE incidence for low-, intermediate- and high-risk tertiles for CARDINAL (joint) and CARDINAL (MoE), with number-at-risk tables.}
\label{fig:latent_survival}
\end{figure*}

A distinctive feature of CARDINAL is its nested family of latent representations, which permits downstream prediction using progressively larger latent prefixes. We evaluated MACE discrimination across the complete Matryoshka depth range, $d\in{1,2,4,\ldots,1024}$ (Fig.~\ref{fig:latent_survival}). The relative depth analysis showed that most of this long-horizon discrimination was retained using substantially smaller representations. CARDINAL (joint) retained approximately 96\% of its depth-sweep peak by $d=64$ and more than 99\% by $d=512$. CARDINAL (MoE) retained approximately 96\% by $d=128$ and nearly 99\% by $d=512$. Similar patterns were observed across the other horizons and labeling conventions, although the optimal depth varied by task. These findings indicate that compact latent prefixes preserve most of CARDINAL's predictive information, while larger representations provide incremental gains in peak discrimination. Complete absolute AUROC and AUPRC results across all depths, horizons, and labeling conventions are provided in Supplementary Information.

\subsection{Risk reclassification}

We next evaluated continuous net reclassification improvement (NRI) using the same held-out predictions as the classification analyses (Table~\ref{tab:primaryevidence}). At 10 years under \textit{withFU}, CARDINAL (joint) showed positive reclassification relative to every non-CARDINAL comparator. NRI was 0.551 (95\% confidence interval [CI], 0.272--0.856) versus structural radiomics, 0.806 (95\% CI, 0.525--1.086) versus CT biomarkers, 0.984 (95\% CI, 0.728--1.264) versus PREVENT ASCVD, 0.933 (95\% CI, 0.655--1.213) versus PREVENT CVD, 0.908 (95\% CI, 0.629--1.188) versus PREVENT BOTH, and 1.059 (95\% CI, 0.801--1.314) versus CAC score. CARDINAL (MoE) also showed positive reclassification, with NRI values ranging from 0.448 versus structural radiomics to 0.802 versus CAC score.

Positive reclassification was also observed at 10 years under \textit{ignoreFU}, where CARDINAL (joint) NRI ranged from 0.073 versus PREVENT CVD to 0.255 versus CAC score. Complete NRI estimates, confidence intervals, and evaluation denominators across all horizons and both labeling conventions are reported in Supplementary Information.

\subsection{Subgroup analyses}

We evaluated 10-year \textit{withFU} performance by sex, age, and CAC burden, with an additional descriptive analysis by race. Across the five fixed models, CARDINAL (joint) AUROC was 0.895 $\pm$ 0.017 in women and 0.847 $\pm$ 0.028 in men, 0.845 $\pm$ 0.037 below age 60 and 0.874 $\pm$ 0.014 at age 60 or older. AUROCs were 0.870 $\pm$ 0.026 for CAC $=0$, 0.862 $\pm$ 0.036 for CAC 1--99, and 0.854 $\pm$ 0.035 for CAC $\geq100$. CARDINAL (joint) was numerically strongest in each of these subgroups except among patients younger than 60 years, for whom structural radiomics achieved AUROC $=0.863 \pm$ 0.014.

Among 130 White patients with 66 events, CARDINAL (joint) achieved AUROC $=0.875 \pm 0.018$, compared with $0.848 \pm 0.014$ for structural radiomics. Among 23 Black or African American patients with 12 events, CARDINAL (joint) achieved AUROC $=0.795 \pm 0.053$, compared with $0.773 \pm 0.030$ for PREVENT ASCVD. Results in the latter subgroup are descriptive given the number of evaluated patients. Subgroup discrimination, comparator identities, and absolute performance differences are summarized in Fig.~\ref{fig:latent_survival}. Complete metrics for every model and subgroup are reported in Supplementary Information.

\subsection{Calibration and clinical utility}

Calibration was assessed for the 10-year \textit{withFU} predictions. Cross-fold isotonic calibration reduced CARDINAL (joint) Brier score from 0.165 $\pm$ 0.015 to 0.148 $\pm$ 0.012 and expected calibration error from 0.117 $\pm$ 0.016 to 0.058 $\pm$ 0.005. Corresponding CARDINAL (MoE) values improved from 0.190 $\pm$ 0.007 to 0.176 $\pm$ 0.006 and from 0.119 $\pm$ 0.017 to 0.068 $\pm$ 0.024. Rank-based discrimination changed minimally. Raw and calibrated metrics and decision-curve analyses are reported in Supplementary Information.

\subsection{Robustness to realistic image perturbations}
We evaluated CARDINAL (joint) at $d=1024$ under controlled intensity, slice-coverage, and spatial perturbations. Across six phenotypes and five models, 30\% intensity reduction decreased mean ICC by 0.008 and increased mean absolute error (MAE) by 4.6\%; 10\% slice dropout decreased ICC by 0.018 and increased MAE by 8.8\%. Thirty-percent spatial masking produced the largest degradation (ICC decrease, 0.127; MAE increase, 50.4\%). Target-level ICC decreases and robustness results are shown in Supplementary Information.

\FloatBarrier

\section{Discussion}


In this study, we developed and evaluated CARDINAL, a clinically grounded
latent representation framework for predicting MACE directly from routine
electrocardiogram-gated non-contrast cardiac CT. The central finding is that a
unimodal image-derived representation can recover clinically meaningful
cardiothoracic phenotypes and support cardiovascular risk prediction beyond
established clinical risk scores, CAC score, and engineered CT features. The
clearest advantage was observed in the 10-year \textit{withFU} setting, where
CARDINAL (joint) achieved the strongest overall long-horizon performance. Its
AUROC was 4.0--21.3 percentage points higher and its AUPRC was
6.4--24.5 percentage points higher than the non-CARDINAL comparators. This
advantage extended to survival discrimination, with a C-index that was
0.066--0.131 higher than the non-CARDINAL comparators and a high-versus-low
predicted-risk tertile HR of 10.78, compared with 2.82--5.92 for the
conventional clinical and engineered imaging approaches. Continuous NRI was
positive against every non-CARDINAL comparator, ranging from 0.551 versus
structural radiomics to 1.059 versus CAC score, and the raw Brier score was
5.2--30.4\% lower. CARDINAL (joint) was also numerically favorable in most
prespecified sex, age, and CAC subgroups, as well as in the descriptive race analysis, including AUROC differences of 2.7 percentage points among White patients and 2.2 percentage points among Black or African American patients relative to the strongest comparator. Structural radiomics was 1.8 percentage points higher among patients younger than 60 years, and the small Black or African American subgroup precludes comparative or fairness conclusions.

The pattern of results is important. CARDINAL did not outperform every baseline in every setting. Gains were strongest in the longer-horizons, especially the follow-up-restricted analyses, and they were more modest under the broader \textit{ignoreFU} convention and at shorter horizons. This distinction is not incidental. The \textit{withFU} setting better aligns with the intended clinical use of prognostic modeling because it restricts evaluation to patients with observed follow-up through the target horizon or an earlier event. By contrast, the \textit{ignoreFU} setting reflects a larger and more pragmatic cohort but necessarily introduces greater label uncertainty, since some patients treated as non-events may simply not have been observed long enough to declare the outcome absent. The attenuation of CARDINAL’s performance gains under \textit{ignoreFU} is therefore consistent with increased outcome noise rather than evidence against the underlying representation. Viewed this way, the stronger \textit{withFU} findings support the argument that CARDINAL is identifying real prognostic signal that becomes diluted when follow-up is incompletely observed.


The 10-year \textit{withFU} analysis is the most clinically consequential result. In that setting, the joint CARDINAL representation exceeded PCE, PREVENT, CAC, and the CT biomarker baselines in matched-cohort discrimination and also showed the strongest survival concordance and high-versus-low risk-tertile HR. The reclassification analyses were directionally consistent with these findings, with large positive NRI values for the CARDINAL (joint) model relative to each traditional baseline. Calibration and decision-curve results were also favorable in this same setting, suggesting that the gains were not limited to rank-order discrimination alone. Taken together, these findings indicate that the strongest value of CARDINAL is not simply that it can separate cases from controls statistically, but that it may provide a more useful patient-level risk ordering for long-term cardiac decision-making.

A key question is why CT could add information beyond CAC and established risk equations. PCE and PREVENT are clinically useful and well validated, but they summarize risk through a limited set of structured variables, depend on the availability and timing of EHR-derived inputs, and have underperformed on minority populations \cite{goff2014american,khan2024development, rana2016accuracy, defilippis2015calibration}. In routine practice, these variables may be missing, incomplete, or temporally misaligned with the imaging encounter. CAC directly measures calcified plaque burden and remains a powerful imaging biomarker, but it is still a scalar summary of a rich three-dimensional examination \cite{agatston1990quantification,Hecht2017-wc}. CARDINAL is designed around the premise that non-contrast cardiac CT contains broader cardiovascular information than can be captured by either a fixed clinical score or a single scalar imaging measurement. The observed improvement over CAC and the two CT biomarker/radiomic baselines supports this premise. The favorable comparison with the more comprehensive 70-feature structural radiomics model further suggests that useful prognostic information is not fully captured by a predefined set of anatomical measurements.

The auxiliary phenotype recovery results help make this interpretation more defensible. CARDINAL was not trained as an unconstrained black box. Its nested latent space was shaped using anatomy- and calcium-grounded supervision, and the learned representations preserved heart, lung, myocardium, aortic, and calcium-related phenotypes with strong agreement, particularly at larger depths. This matters because, with the help of MRL, it suggests that the model is organizing image information in a clinically anchored way rather than extracting a purely opaque embedding disconnected from recognizable biology. The fact that the same latent space both preserved these phenotypes and supported downstream MACE prediction strengthens the claim that CARDINAL is learning a clinically meaningful cardiothoracic representation rather than merely exploiting spurious correlations.

The MRL structure is also important from a translational perspective. Although the highest long-horizon performance generally occurred at larger dimensions, much of the available predictive signal was retained in substantially smaller latent prefixes. For example, CARDINAL (joint) reached at least 95\% of its maximum 10-year \textit{withFU} AUROC at 64 dimensions, while CARDINAL (MoE) reached this threshold at 128 dimensions. At several shorter-horizon tasks, at least 95\% of peak performance was retained using only two to four dimensions. Relative to an input volume of $32\times224\times224$, containing approximately 1.6 million voxels, the 1024-dimensional representation provides more than 1,500-fold dimensionality reduction, with substantially greater compression at smaller prefixes. This supports a practical deployment model in which a compact representation is computed once, stored efficiently and reused for risk prediction, retrospective investigation, or future multimodal fusion. In other words, CARDINAL is not only a predictive model, but also a reusable imaging representation with considerable dimensionality reduction. That distinction is part of the novelty. The framework is not merely producing one task-specific risk score, but a nested family of clinically grounded representations that can support multiple downstream uses while remaining storage-efficient and operationally tractable.


The comparison between the joint and MoE models adds further nuance. The joint representation was the strongest overall model, especially at 10-year \textit{withFU}, whereas the MoE model was competitive and achieved the highest AUROC at the 3- and 5-year \textit{ignoreFU} horizons. This suggests that both globally integrated and compartment-specific image information contribute to cardiovascular risk, but that for long-horizon prognostication a unified shared latent space may better capture distributed disease burden across the thorax. In other words, the prognostic phenotype appears to be broader than any single organ-specific descriptor, which helps explain why the joint model generally produced the strongest overall results.

The subgroup analyses are also informative. In the 10-year \textit{withFU} setting, CARDINAL remained favorable across sex, age, race, and CAC strata, with especially notable gains in women, older patients, Black or African American people, and the CAC 1–99 subgroup. The latter may be particularly relevant clinically because intermediate calcium burden is often where treatment decisions are most uncertain. At the same time, subgroup performance was not uniformly improved in every setting, especially under \textit{ignoreFU}. That heterogeneity is important to acknowledge. It argues against overclaiming universal superiority, but it does not weaken the broader translational case. Instead, it suggests a more realistic conclusion: CARDINAL appears most valuable in settings where long-term follow-up is meaningful, event labeling is reliable, and incremental risk signal beyond conventional markers is most needed. Notably, these results were obtained without incorporating clinical variables such as age, sex, race, or blood pressure into the model. This suggests that the image-derived representation captures independent prognostic information, enabling the model to identify high-risk subgroups using imaging data alone, without explicit knowledge of baseline patient characteristics.

The study cohort also supports the practical relevance of these findings to
real-world U.S. imaging practice. Patients were drawn from 11 sites within a
large healthcare system, included a demographically diverse population and were
imaged using scanners from five recorded manufacturer labels. This multisite,
multivendor design captures meaningful variation in patient characteristics,
clinical workflows, and image acquisition that is often absent from
single-center or single-vendor studies. The cohort's heterogeneity improves its relevance to the broader U.S. clinical population.

Calibration and robustness provide separate information from discrimination. Cross-fold isotonic calibration improved Brier score and expected calibration error with little change in rank metrics, but remained exploratory without an independent calibration cohort \cite{Van_Calster2019-wk,vanCalster2016hierarchy,steyerberg2010performance}. Modest intensity reduction and slice dropout produced limited phenotype-recovery degradation, whereas spatial masking produced the largest decline. This is consistent with a representation that tolerates modest acquisition and coverage changes but still depends on preserved anatomy. Motion, truncation, reconstruction-kernel shifts, and external scanner variation require dedicated validation.

From a deployment perspective, the practical appeal of CARDINAL lies in its ability to extract additional cardiovascular risk information from scans that are already being acquired. Routine non-contrast cardiac CT is performed at large scale for diverse clinical reasons, including thoracic evaluation and lung cancer screening. An approach that derives opportunistic cardiovascular risk phenotypes from these existing scans could expand preventive assessment without additional radiation, acquisition time, or specialized protocols. This is especially attractive in real-world environments where structured risk-factor data may be incomplete. The fact that CARDINAL is image-only at inference reduces dependence on EHR completeness, avoids re-entry of clinical variables, and creates a pathway for scalable retrospective and prospective implementation.

The novelty of this work therefore rests on several linked contributions. First, CARDINAL learns from routine non-contrast cardiac CT rather than dedicated cardiac imaging alone. Second, it does so through a clinically grounded latent representation rather than through a purely end-to-end black-box classifier or a fixed handcrafted feature pipeline. Third, it produces nested, compact embeddings that preserve anatomically meaningful information while supporting downstream cardiovascular risk prediction. Fourth, the framework showed its strongest benefit in comparisons against clinically relevant baselines, including both conventional risk equations and imaging-based baselines. Together, these features distinguish CARDINAL from scalar biomarker approaches, radiomics-only approaches, and task-specific black-box classifiers as the first presented image-only approach for cardiovascular risk assessment.

Future work should emphasize external validation, deeper analysis of discordant cases relative to CAC and conventional scores, and multimodal extension. CARDINAL is especially well positioned for multimodal fusion because it converts the CT into a compact representation that can be combined with laboratory values, medications, electrocardiogram features, and/or clinical text. Prospective impact studies should also test whether CARDINAL-guided reclassification changes lipid-lowering or other preventive treatment, improves adherence or referral patterns, and ultimately reduces cardiovascular events. More broadly, the framework suggests a different way to think about routine CT in cardiovascular prevention: not merely as a source of one or two opportunistic biomarkers, but as a reusable phenotypic record from which clinically grounded latent representations can be learned and deployed.

\section{Methods}
\label{methods}
\subsection{Study design}
This retrospective cohort study developed and evaluated CT-based models for predicting MACE from non-contrast cardiac CT. The primary objective was to determine whether a clinically grounded latent representation derived from CT imaging improves risk prediction relative to established clinical and imaging-based approaches.

The analysis pipeline consisted of: (1) representation learning from CT imaging without access to outcome labels, (2) downstream prediction of MACE using CT representations, and (3) evaluation across discrimination, calibration, reclassification, and survival modeling. All modeling and evaluation steps were performed using strictly separated patient-level splits to prevent information leakage.

The study was conducted in accordance with Transparent Reporting of a multivariable prediction model for Individual Prognosis Or Diagnosis plus Artificial Intelligence (TRIPOD+AI) \cite{collins2015tripod,Collins2024-gz} and interpreted using Prediction model Risk Of Bias ASsessment Tool plus Artificial Intelligence (PROBAST+AI) principles \cite{wolff2019probast,Moons2025-wh}.

\subsection{Ethics}
This study was approved by the Emory University Institutional Review Board, which waived the requirement for informed consent because of the retrospective design and use of de-identified clinical and imaging data. All procedures were conducted in accordance with relevant ethical guidelines and regulations.

\subsection{Study population}
We identified adults who underwent electrocardiogram-gated non-contrast cardiac CT for clinical evaluation or CAC scoring between 2010 and 2023 across 11 Emory Healthcare-affiliated sites in Atlanta, Georgia. Examinations were acquired using scanners from Philips, Siemens, General Electric, Toshiba, and Canon Medical Systems. Imaging data were linked to longitudinal EHR data.

History of ASCVD and imaging indication were determined using historical ICD and CPT codes together with CT order information.

Each patient contributed one index CT examination, defined as the earliest eligible scan. Patients with multiple examinations were restricted to this index study before model development or data partitioning. Examinations were excluded for inadequate image quality or cardiothoracic coverage, preprocessing incompatibility, or insufficient data linkage. After index-examination selection and eligibility assessment, 17,659 patients comprised the final analytic cohort. Study size was determined by all eligible patients in the retrospective source cohort; no formal prospective sample-size calculation was performed.

The final cohort was partitioned at the patient level into training (70\%), validation (10\%), and held-out test (20\%) sets. Splits were stratified by MACE outcome and baseline calcium burden to preserve outcome prevalence and disease-severity distributions across partitions. No patient appeared in more than one partition at any stage of model development or evaluation.

\subsection{Outcome definition}
MACE were defined as the occurrence of stroke, myocardial infarction, percutaneous coronary intervention, coronary artery bypass grafting occurring more than 90 days after the index CT, or all-cause mortality.

Events were identified using diagnosis and procedural codes supplemented by mortality data. To validate outcome definitions, clinicians manually reviewed over 10\% of the cohort, including representation from each event category and non-event cases. Outcome extraction and review were completed independently of downstream model predictions.

For horizon-based classification, binary labels were defined at 1-, 3-, 5-, and 10-year time horizons. A case was labeled positive if an event occurred within the specified time window.

To account for incomplete follow-up, two labeling strategies were used:
\begin{itemize}
\item \textit{withFU}: only patients with documented follow-up through the specified horizon or an earlier event were included; negative labels therefore correspond to observed event-free follow-up.
\item \textit{ignoreFU}: absence of a recorded event by the specified horizon was treated as a negative label regardless of follow-up completeness.
\end{itemize}


For survival analyses, time-to-event was defined as the duration from the index CT to first MACE or last documented follow-up. Patients without an observed MACE were right-censored at their last documented follow-up. Follow-up beyond 10 years was administratively censored at 10 years. Survival eligibility was determined from valid event or censoring times and was independent of the horizon-specific \textit{withFU} and \textit{ignoreFU} classification labels.

\subsection{Clinical and imaging baselines}
PREVENT ASCVD \cite{khan2024development} was the primary clinical comparator and was calculated according to its published specification using EHR-derived variables recorded at or during the 6 months preceding the index CT. Age was defined at the index date. Inputs comprised age, sex, body mass index, systolic blood pressure, total and high-density lipoprotein cholesterol, estimated glomerular filtration rate, antihypertensive treatment, lipid-lowering treatment, diabetes, and current smoking. Historical clinical risk-equation results \cite{goff2014american} were calculated for supplementary comparison only. Cases missing required inputs were excluded from analysis of the corresponding clinical equation; variable-specific available-case counts are reported in Supplementary Table 1.

CAC burden was quantified using the Agatston score \cite{agatston1990quantification} using a validated automated pipeline \cite{Winkel2022-ml, Martin2020-df}.

A six-feature CT biomarker baseline used segmentation-derived heart, myocardium, lung and aorta volumes, CAC volume, and CAC score. Segmentations were derived using TotalSegmentator \cite{wasserthal2023totalsegmentator} and the study calcium pipeline. The radiomics benchmark was a distinct, additional 70-feature structural comparator containing fourteen three-dimensional PyRadiomics shape features for each of the aorta, left atrium, left ventricle, right atrium, and right ventricle.

For each horizon and labeling convention, all displayed models were evaluated on the same held-out patients with available predictions. The displayed models were CARDINAL (joint), CARDINAL (MoE), structural radiomics, CT biomarkers, PREVENT ASCVD, PREVENT CVD, PREVENT BOTH, and CAC score. PCE results were retained in Supplementary Information. Classification metrics and NRI were computed using paired patient-level predictions. Survival metrics were computed in patients with valid event or censoring times, with follow-up administratively censored at 10 years. Evaluation denominators and event counts are reported with the corresponding analyses.

\subsection{CT preprocessing}
CT volumes were converted to Neuroimaging Informatics Technology Initiative (NIfTI) format and processed using a standardized pipeline. Images were reoriented to a canonical anatomical axis, intensity-normalized, and resampled to a fixed spatial resolution.

Volumes were resized to a consistent grid to enable uniform model input. A deterministic slice-selection strategy was applied to ensure consistent inclusion of cardiothoracic structures while preserving global thoracic context. The same preprocessing and quality-control criteria were applied across demographic groups.

Segmentation masks derived from TotalSegmentator \cite{wasserthal2023totalsegmentator} and CAC annotations \cite{Follmer2024-zh} were used both to guide preprocessing and to compute auxiliary anatomical targets used during representation learning.

\subsection{Model development}
CARDINAL is a three-dimensional representation learning framework that encodes non-contrast CT into a compact latent representation while preserving clinically meaningful anatomical and calcium-related information. The encoder backbone is a Swin Transformer \cite{liu2021videoswintransformer}. MRL was used to construct a nested latent space in which progressively larger prefixes retain additional information. Representation architecture and training procedures were defined before held-out test evaluation. All completed latent dimensions and downstream candidates are retained in Supplementary Information.

Importantly, representation learning was performed without access to MACE outcomes. Instead, supervision was provided using CT-derived anatomical and calcium-related auxiliary targets, ensuring that the learned representation is clinically grounded while preventing label leakage from downstream prediction tasks.

\subsection{Joint and mixture-of-experts (MoE) representations}
We evaluated three representation outputs. In the joint formulation, a single encoder was trained to predict all six anatomical and calcium-related targets, producing one unified nested latent representation. In the single-expert formulation, six separate encoders were trained, one each for aorta volume, heart volume, lung volume, myocardium volume, CAC volume, and CAC score.

In the fused/MoE formulation, the six 1024-dimensional expert embeddings were concatenated and passed through a learned multilayer fusion module to produce one 1024-dimensional fused nested representation comparable with the joint representation. The fusion run did not substitute a single expert or average expert-level phenotype metrics. All encoders and the fusion model were trained without access to MACE outcomes and were frozen before downstream event modeling.

These approaches enable comparison between a single shared representation and a modular composition of specialized representations.

\subsection{Downstream modeling}
After representation learning, encoder weights were frozen and applied to all CT studies to extract latent embeddings. Downstream models were trained using only the training set, with hyperparameter tuning performed on the validation set. Model selection was conducted independently within each training fold using predefined performance criteria (largest AUROC followed by AUPRC). To avoid overfitting and selection bias, final reported results correspond to the best-performing model configuration identified on validation data and then evaluated once on the held-out test set. Nested cross-validation and bootstrap resampling were used to estimate variability in performance metrics. All model development decisions were made without access to test set outcomes.

Downstream model selection was performed over a predefined set of model classes, including FT-Transformer (Feature Tokenizer Transformer), multilayer perceptron, XGBoost, CatBoost, and random forest, with deterministic and Monte Carlo inference variants to the deep learning models. For survival modeling, penalized Cox proportional hazards models with elastic net regularization were used. Model comparison was limited to a predefined set of candidate architectures to reduce multiplicity. Hyperparameters for each model class were tuned using the validation set within each training fold using grid search within fixed parameter ranges specified a priori on a Dell Pro Max T2 with an NVIDIA RTX PRO 6000 Blackwell GPU. 

\subsection{Survival modeling}





Time-to-event modeling was performed using penalized Cox proportional hazards models with elastic-net regularization. One model was trained for each of five fixed-fold training and validation partitions and evaluated on the same held-out test partition.

The survival cohort included patients with a valid time from the index CT to first MACE or last documented follow-up. Patients without an observed event were right-censored at last follow-up, and follow-up beyond 10 years was administratively censored at 10 years. The same fitted Cox model generated predicted risks at 1, 3, 5, and 10 years.

Latent representations and baseline model inputs were used as Cox covariates. Harrell's C-index quantified overall survival discrimination. Brier scores at 1, 3, 5, and 10 years were calculated as secondary metrics and are reported in Supplementary Information.

For risk stratification, patient-level risks were averaged across the five fixed models and divided into model-specific tertiles. High-versus-low tertile hazard ratios and two-sided log-rank tests quantified risk separation, and cumulative incidence was displayed as one minus the Kaplan--Meier estimate \cite{cox,harrell1996multivariable,royston2013external}.

\subsection{Evaluation metrics}
For classification, primary metrics included the area under the receiver operating characteristic curve (AUROC) and area under the precision--recall curve (AUPRC). Additional metrics included sensitivity, specificity, F1 score, Brier score, calibration slope and intercept, expected calibration error (ECE), and decision-curve analysis \cite{steyerberg2010performance,vickers2006decision,vanCalster2016hierarchy}.

Risk reclassification was assessed using continuous, category-free net reclassification improvement (NRI). NRI quantified the extent to which CARDINAL assigned higher predicted risk to patients with events and lower predicted risk to patients without events relative to each comparator. Comparisons used the same paired patient-level predictions as the corresponding classification analysis. Positive NRI values favor CARDINAL. Confidence intervals were estimated using patient-level, event-stratified nonparametric bootstrap resampling.

For survival analysis, performance was evaluated using C-index, high-versus-low predicted-risk tertile HRs, and cumulative-incidence separation.

Raw calibration, decision-curve analysis, and reclassification were computed on held-out test predictions. Platt and isotonic calibrators were fitted across four prediction folds and applied to the fifth. Platt scaling is monotone and therefore preserves rank-based discrimination, whereas isotonic score ties can produce small numerical changes in rank metrics.

\subsection{Subgroup analyses}

Prespecified subgroup analyses were performed on held-out test predictions by sex (women, men), age ($<60$ years, $\geq60$ years), race (White, Black or African American), and baseline CAC burden (CAC $=0$, CAC 1--99, CAC $\geq100$). Within each labeling convention and horizon, all models were evaluated on the same patients within each subgroup. AUROC was computed for CARDINAL (joint), CARDINAL (MoE), and each available baseline. The main figure reports the strongest non-joint comparator within each subgroup. Full AUROC, AUPRC, sensitivity, specificity, precision, accuracy, F1, Brier score, and ECE results are provided in Supplementary Information.

\subsection{Robustness analysis}
Robustness was assessed for the joint representation at latent dimension 1024 on the held-out test split across five folds and six phenotype targets. Retained perturbations were intensity scaling, slice dropout, and spatial masking, selected as acquisition or coverage stresses with direct imaging analogues. Gaussian-noise perturbations were excluded from the reported analysis. Degradation was summarized relative to the clean test baseline using MAE increase, $R^2$ drop, ICC drop, and the proportion of targets exceeding a 25\% true-value error threshold. Full target- and fold-level results are included in Supplementary Information.

\subsection{Statistical analyses}

For classification and survival modeling, metrics were computed for each of five independently trained fixed-fold models on the held-out test patients and are reported as mean $\pm$ standard deviation.

For risk reclassification, continuous NRI was estimated separately for each CARDINAL comparator pair using the corresponding paired held-out predictions. Patients and their paired predictions from both models were resampled using event-stratified nonparametric bootstrap resampling to estimate 95\% confidence intervals. NRI analyses were category-free and did not depend on prespecified clinical risk thresholds.

Pairwise AUROC comparisons used two-sided DeLong tests on paired patient-level held-out predictions \cite{delong1988comparing}. Complementary exact two-sided Wilcoxon signed-rank tests compared five paired model-level metric values. Survival risk groups were compared using two-sided log-rank tests and Cox model contrasts. Statistical significance was defined as $\alpha=0.05$.

All analyses were conducted using Python with standard scientific computing libraries.

\subsection{Use of generative artificial intelligence}
OpenAI was used during manuscript preparation for language editing. The authors reviewed the source data, analysis code, numerical results, and final text and take responsibility for the work.

\backmatter
\bmhead{Supplementary information}
Supplementary information provides additional classification, reclassification, survival, calibration, subgroup, phenotype-recovery, and robustness results.

\newpage

\section{Declarations}
\paragraph{Ethics approval and consent to participate.} The retrospective study was conducted under institutional review board oversight with waiver of informed consent.
\paragraph{Competing interests.} A provisional patent application related to this work has been filed.
\paragraph{Funding.} Dr. De Cecco received research funding from Siemens, Cleerly, Elucid, and Pfizer.
\paragraph{Correspondence.}
Correspondence and requests for materials should be addressed to Dr. Ali Adibi
(\href{mailto:ali.adibi@ece.gatech.edu}{ali.adibi@ece.gatech.edu}).

\newpage
\bibliography{bibliography}
\end{document}